\documentclass{ws-ijmpe}
\usepackage{graphicx}
\usepackage{bm}

\begin{document}
\title{Hydrodynamic Scaling Analysis of Nuclear Fusion \\driven by ultra-intense laser-plasma interactions}
\author{Sachie Kimura\footnote{Current address: Department of Physics, Universita' degli Studi di Milano, via Celoria 16, 20133 Milano, Italy}}
\address{INFN-LNS, via Santa Sofia, 62, 95123 Catania, Italy}
\author{Aldo Bonasera}
\address{INFN-LNS, via Santa Sofia, 62, 95123 Catania, Italy and \\
Cyclotron Institute, Texas A\&M University, College Station TX 77843-3366, USA}

\maketitle

\begin{abstract}
We discuss scaling laws of fusion yields generated by laser-plasma interactions.
The yields are found to scale as a function of the laser power.
The origin of the scaling law in the laser driven fusion yield is derived in terms of hydrodynamic scaling.
We point out that the scaling properties can be attributed to the laser power dependence of three terms: 
the reaction rate, the density of the plasma and the projected range of the
plasma particle in the target medium.
The resulting scaling relations have a predictive power 
that enables estimating the fusion yield for a nuclear reaction which has not been investigated 
by means of the laser accelerated ion beams.   
\end{abstract}

\section{Introduction}
Ultraintense laser pulses can generate high energy protons when irradiating a thin foil target~\cite{PhysRevLett.84.670,PhysRevLett.85.2945,foord,spencer}.
The proton beams generated by the laser irradiations 
have various potential applications~\cite{ledinghamrev} in science, engineering and medical imaging~\cite{spencer2,kb-dirlp}.    
The laser pulse parameters, i.e.,  pulse energy, peak intensity and pulse duration,
are, however, currently not optimized for practical applications.
The determination of scaling laws in both the characteristics of the accelerated ion beams itself and in the resulting fusion yield
is crucial for such an optimization. 
In this respect, we restrict our study on the scaling laws of fusion yields derived empirically from the currently available experimental 
data. We do not consider the acceleration mechanism of rediation pressure acceleration~\cite{hofmann} 
that could be important to obtain higher energy ion beams for the 
oncological proton(or hadron)-therapy~\cite{pla}, because such a mechanism is not feasible at current technology. 
Clarifying the scaling laws of the fusion yield 
promises to extend the result
even to estimate fusion yield for not well-investigated nuclear reactions by means of the ultrahigh-intensity laser-matter interactions, 
e. g., aneutronic fusion reactions~\cite{belyaev:026406,kb-com}
and deuteron-induced reactions~\cite{kb-dirlp,fujimoto,petrov,willingale}.
Especially deuteron-induced reactions are expected to induce enough yields for  
the radioactive isotope production for PET diagnostics using laser-induced ions.
For the purpose of the radioactive isotope production for PET diagnostics, 
monochromatic ion beam with an energy as high as 200 MeV for oncological hadron-therapy 
is not necessary. 

%
On this issue, scaling laws of the maximum energy and conversion efficiency of laser-accelerated protons from thin foil targets have been studied 
~\cite{fuchs_nature,robson,maxpresc}.
When a laser of intensity higher than 10$^{19}$~W/cm$^2$ is irradiated on an aluminum foil target,
a simple power scaling of the form $E_{pmax}=a \times I^b$ with $b=0.5 \pm 0.1$ is found~\cite{robson,spencer2}, 
where $E_{pmax}$ and $I$ denote the maximum proton energy and the laser pulse intensity, respectively.
More in general the maximum proton energy is derived also as a function of the laser power~\cite{maxpresc}.
The dependence of the conversion efficiency on the laser energy is found to be linear, as it is shown in Fig. 2 in~\cite{robson}.

The scaling feature of the accelerated protons is obviously reflected in the yield of the laser-induced nuclear reactions. 
The neutron yield in deuterium cluster explosions is found to scale as a function of the laser pulse energy~($E_L$)~\cite{PhysRevA.70.053201,utanr}
and it follows a quadratic power law dependence.
The other examples are the $^{63}$Zn yield~\cite{robson} and the $^{11}$C yield~\cite{ledingham}  
for the reactions $^{63}$Cu($p,n$)$^{63}$Zn and $^{11}$B($p,n$)$^{11}$C, respectively. 
Both reactions are obtained with proton beams generated at
the VULCAN laser of the Rutherford Appleton Laboratory~(RAL).
The isotopes $^{63}$Zn and $^{11}$C decay by emitting a positron which annihilates an electron producing a pair of photons 
with an energy of 0.51 MeV.  The $^{63}$Zn and $^{11}$C activities are determined by detecting the 0.51 MeV $\gamma$-rays. 
The $^{63}$Zn activity shows again a quadratic power dependence on the laser pulse energy (Fig. 3 in~\cite{robson}).

The above mentioned experiments use different types of primary targets: solid-aluminum foils and 
deuterated clusters. The mechanisms of the ion acceleration from these targets are thought to be
not unique, i.e., in the case of the foil targets, the target normal sheath acceleration~(TNSA) mechanism~\cite{hatchett,passoni} is responsible 
for the ion acceleration. Within this mechanism electrons on the target laser-irradiated surface are accelerated by the laser pulse 
in the direction of the laser propagation. The electrons pass through the target, ionizing the surface contaminants, 
and form a sheath on the rear surface. The charge separation due to the motion of electrons  causes the electrostatic field which 
accelerates the protons in the surface contaminants.
While, in the case of deuterated cluster target irradiation, the deuteron-acceleration is attributed to the bound-electron expulsions 
from the clusters by laser-irradiations~\cite{natureDitmire}.    
Nevertheless the obtained experimental data on the fusion yield show a similar quadratic dependence on the laser pulse energy.    
Two characteristic commons to those 
experiments are the spectra of the accelerated ions which have a Maxwellian-like
shape~\cite{ledingham,td74,td88} and the fact that the fusion reaction occurs in the beam-target mechanism~\cite{td88,kb-comnusp}. 
These data have not been investigated comprehensively, because of the apparent difference in the ion acceleration mechanisms. 
In this paper we analyze these data assuming hydrodynamic scaling~\cite{PhysRevLett.59.630,csernai} and show that 
the scaling laws in fusion yields can be explained inclusively.
For this purpose we reformulate the scaling of proton acceleration found in the maximum proton energy and 
the conversion efficiency~\cite{robson,fuchs_nature,maxpresc}. These two features are related with the temperature, i.e., the width of the Maxwellian-like spectra,  
of the accelerated ions and the total number of the accelerated ions: both scale 
as functions of the laser power. 
These two parameters are, indeed, more fundamental than the maximum proton energy in the sense of
characterizing the energy spectra of the accelerated ions and 
estimating the nuclear reaction rate, as is described in~\ref{sec:rrppA}.  
In this context we stress the importance of determining the plasma temperature for the yield evaluation.  


 \bigskip

\section{Scaling of fusion yields}
\label{sec:sfy}

\begin{figure}[]
  \centering
  \includegraphics[height=.36\textheight, bb=0 0 846 594]{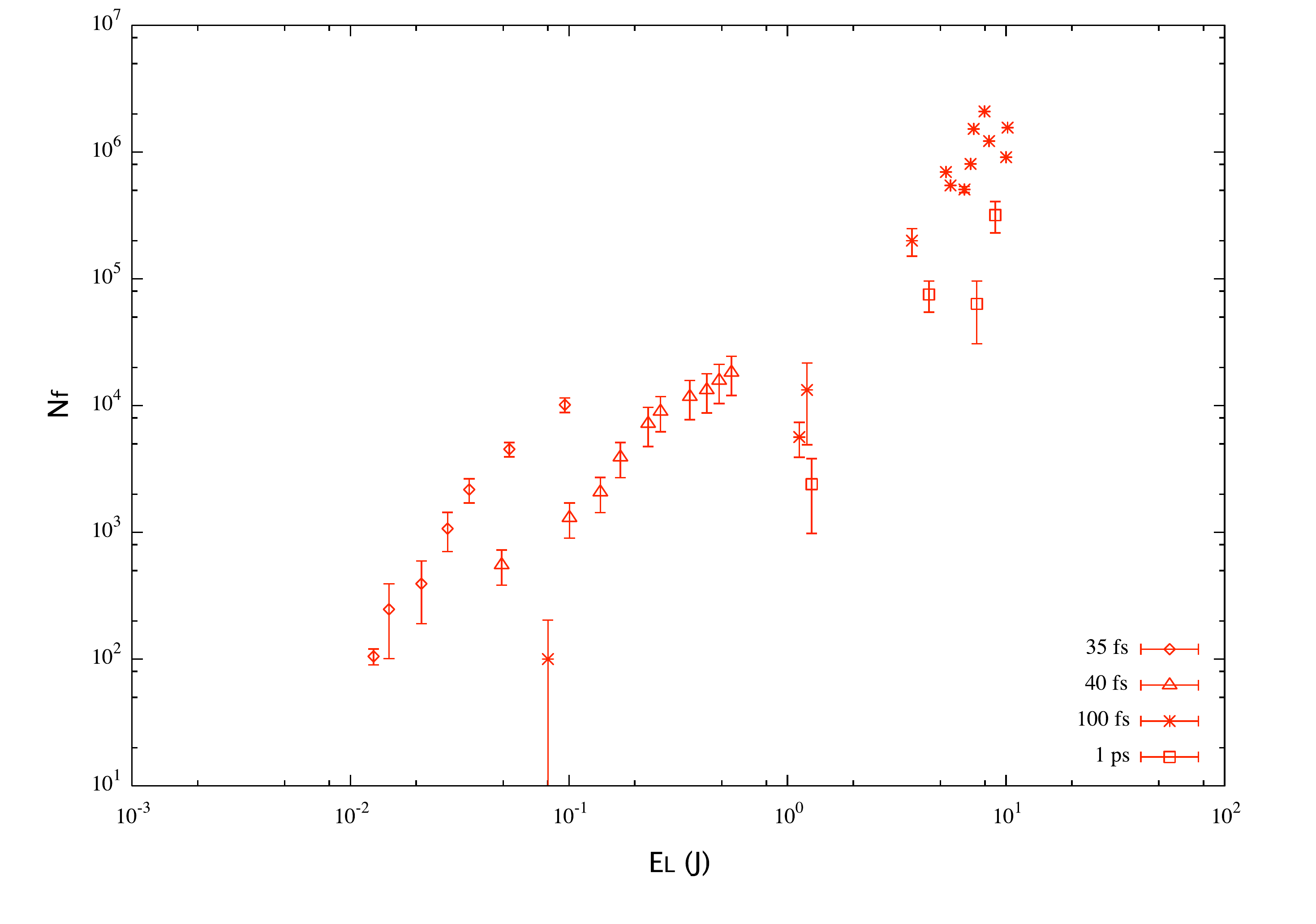}
  \includegraphics[height=.36\textheight, bb=0 0 846 594]{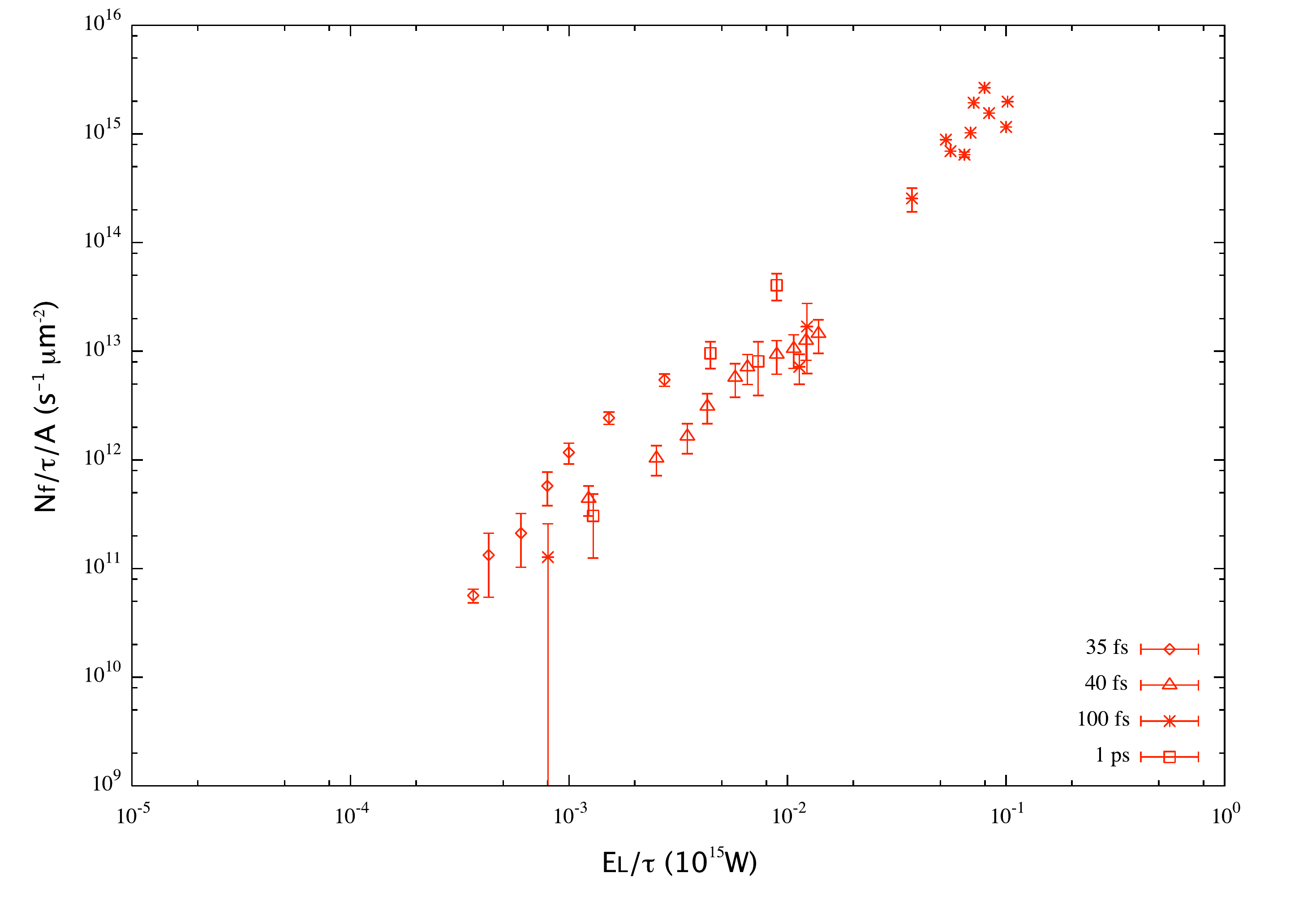}    
  \caption{(Top panel)~Fusion yield of the reactions D(d,n)$^3$He by the cluster explosions as a function of the laser pulse energy. The experimental data are retrieved from Ref. [19] (pulse duration 35 fs~(diamonds)) and with THOR laser at University of Texas at Austin(pulse duration 40 fs~(triangles))~[25] and with JanUSP~(Calisto) laser at LLNL~(pulse duration 100 fs~(stars) and 1 ps (squares))~[18].   (Bottom panel)~Scaled $N_f/\tau/A$ as a function of the laser power. The symbols for the fusion yield data are the same of the top panel. }
  \label{fig:yield}
\end{figure}

We begin with the scaling features of the neutron yield from deuterium cluster explosions,
which indeed motivated us studying the scaling of the yield from laser-induced nuclear-reactions.  
The experiments have been conducted by the group of the university of Texas at Austin.
The neutron yield from deuterium cluster explosions scales as a function of the laser pulse energy~($E_L$)~\cite{PhysRevA.70.053201}:
the yield follows a quadratic power law dependence on the pulse energy. 
The same group has extended the yield measurement using different lasers with different laser parameters
~\cite{physPlas7}. The results are summarized~\cite{utanr} and they are reproduced in the top panel of  
Fig.~\ref{fig:yield}. The yield ($N_f$) shows not only the quadratic dependence but also a clear dependence on the pulse duration~($\tau$).
By reanalyzing these data, we found that all points fall in a line of a quadratic power law, 
if $(N_f \tau /A)$, where $A$ is the laser spot size, is plotted as a function of the laser pulse energy.
In other words, 
\begin{equation}
	\label{eq:nf0}
	N_f \tau /A \propto E_L^2.
\end{equation}
Dividing Eq.~(\ref{eq:nf0}) by $\tau^2$, the power law dependence of the yield becomes:   
\begin{equation}
	\label{eq:nf00}
 	\frac{N_f}{\tau A}  \propto \left(\frac{E_L}{\tau}\right)^2.
\end{equation}
This equation is suggesting that the quantity $(N_f/\tau/A)$ has a power law scaling with an exponent of 2. 
The quantity $(N_f/\tau/A)$  is shown in the bottom panel of Fig.~\ref{fig:yield} where
the pulse-duration dependence seen in the top panel is disappeared. 

\begin{figure}
  \includegraphics[height=.38\textheight, bb=0  0 846 594]{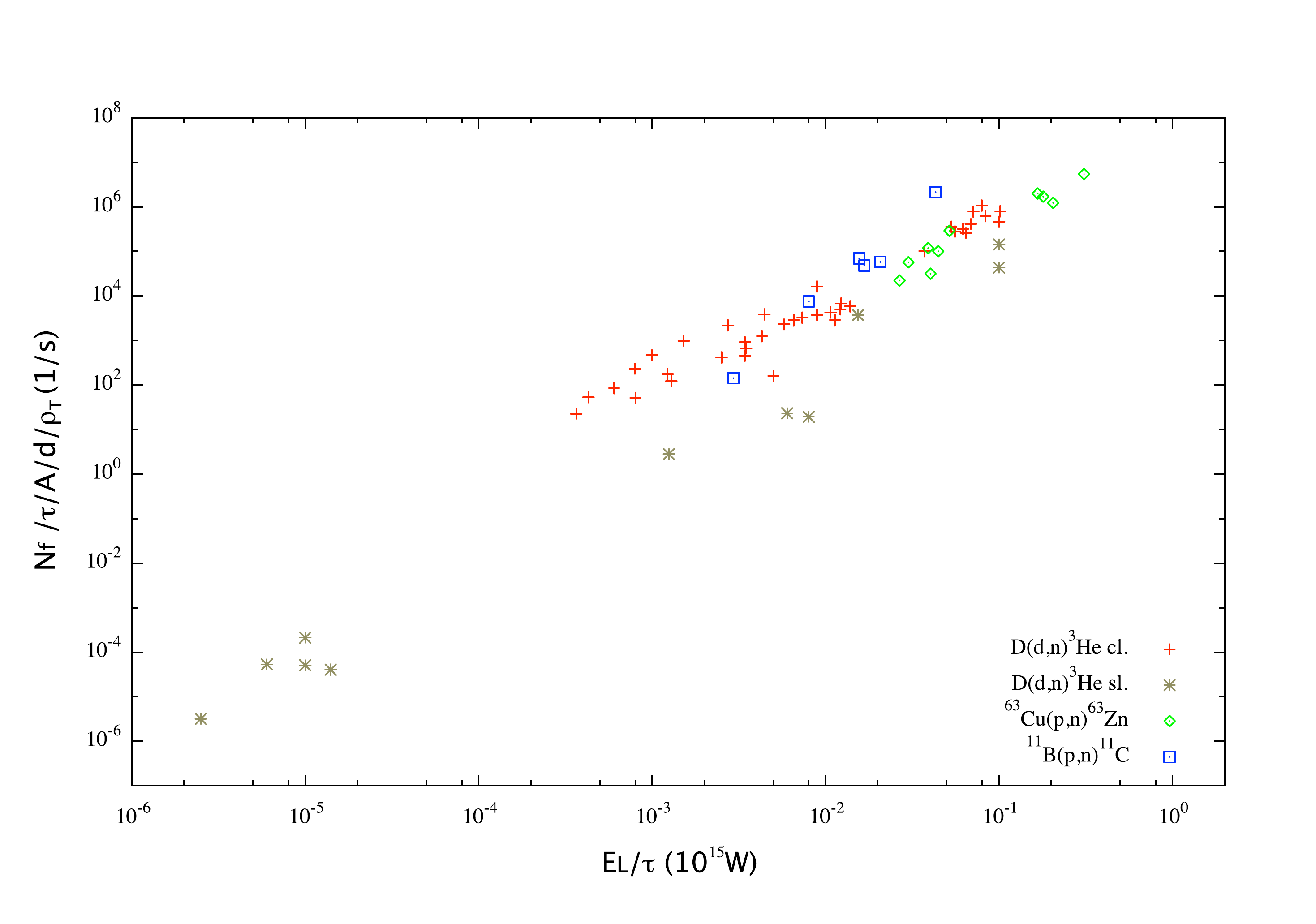}

  \caption{ \label{fig:T} The $\Pi$ for the reaction D($d,n$)$^3$He with cluster targets~(plusses) and 
  with	solid targets~(stars), for the reaction $^{63}$Cu($p.n$)$^{63}$Zn with copper stack~(diamonds) and for the reaction
  $^{11}$B($p,n$)$^{11}$C with boron slab target~(squares) at laser facilities worldwide.}
\end{figure}
\begin{figure}
  \includegraphics[height=.38\textheight, bb=0  0 846 594]{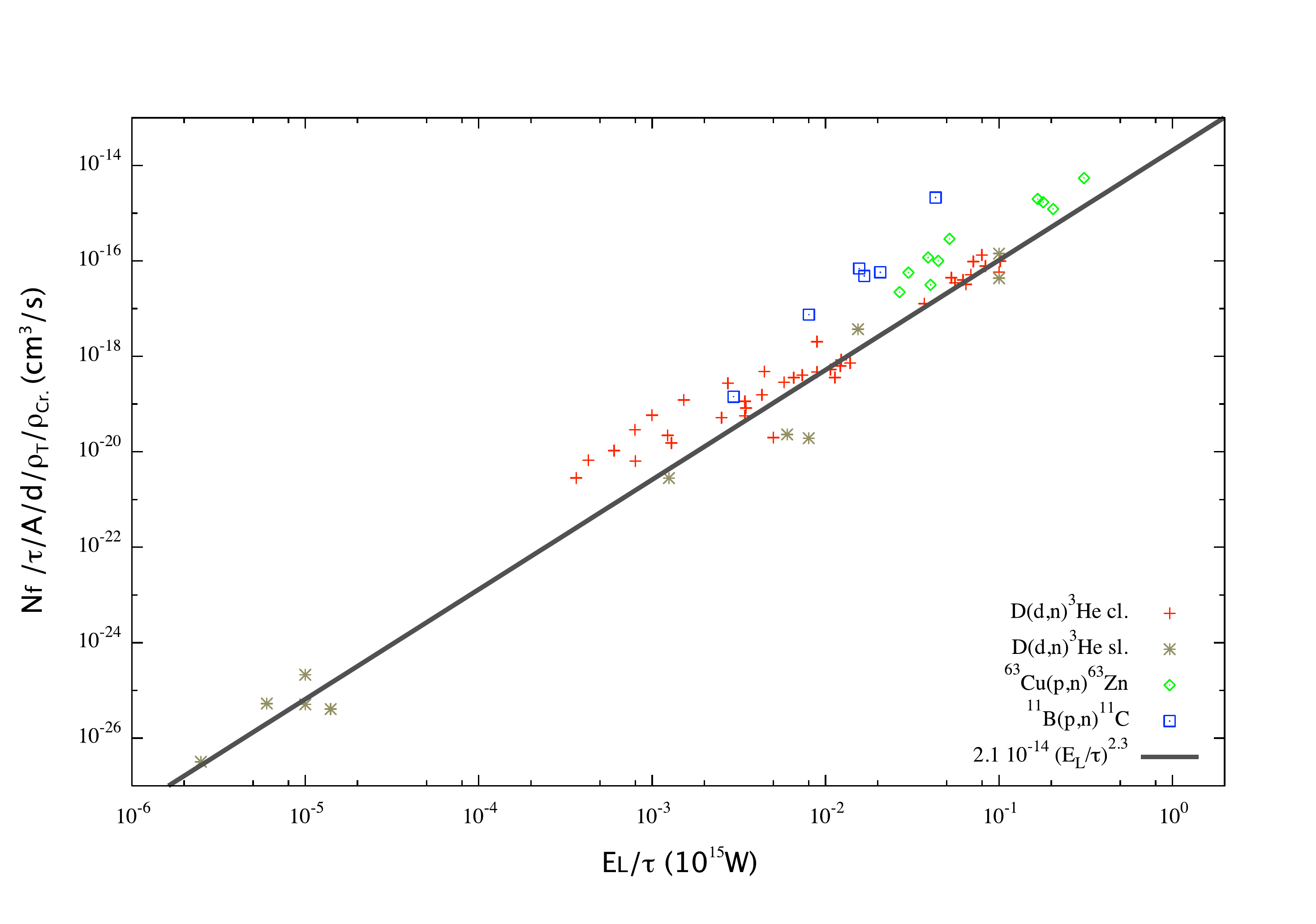}

  \caption{ \label{fig:Cr} The $\Pi$ divided by the plasma critical density~($\rho_{cr}$) for the reactions in Fig.~\ref{fig:T}.    
  The symbols are the same of Fig.~\ref{fig:T}.  
  The cluster target data are divided by a factor 8, as it is explained in the text. The curve is a fitting to all the data for 
  the reaction D($d,n$)$^3$He.}
\end{figure}

The scaling features found in the above experimental data of neutron yield are observed in the experimental data by other groups. 
We analyze a wider range of experimental data by different groups worldwide together with the data by the group of the university of Texas at Austin. 
The neutron yields from three nuclear reactions
D($d,n$)$^3$He,  $^{11}$B($p,n$)$^{11}$C and $^{63}$Cu($p,n$)$^{63}$Zn are analyzed.
Experimentally the data are obtained by 
laser-pulse irradiations on primary targets of 
plastic deuterium (CD), deuterium clusters, boron slabs and copper foils.
In  Tab.~\ref{tab:1} we summarize the laser parameters and the observed neutron yields from the reaction D($d,n$)$^3$He for various experiments.
\begin{table*}[htdp]
\caption{Laser parameters, targets and the observed neutron yield for the reaction D($d,n$)$^3$He at various laser facilities. $^a$The laser pulse energy is not specified in the paper and replaced by the maximum laser pulse energy available from the laser system. $^b$The target thickness is not specified, but replaced by the absorption length in Yamanaka et al. }
\begin{center}
{\footnotesize
\begin{tabular}{r r @{.} l r r@{.}l crr}
\hline\hline
\multicolumn{1}{c}{Reference} & \multicolumn{2}{r}{Pulse energy} & length& \multicolumn{2}{l}{Focal spot}            & Target thick- & \multicolumn{1}{c}{material} & \multicolumn{1}{c}{neutron yield} \\
                   & \multicolumn{2}{l}{(J)}                   &  \multicolumn{1}{c}{(fs)} &    \multicolumn{2}{l}{FWHM($\mu$m)} & ness($\mu$m)           &                &          \\
\hline
Pretzler et al.~\cite{PhysRevE.58.1165}    & 0&2 & 160. & 4&5 &  200. & C$_2$D$_4$ solid & 1.4$\times$10$^2$ \\
Hilscher et al.~\cite{PhysRevE.64.016414} & 0&3 & 50. & 15& & 200. & CD$_2$ solid &  4.$\times$10$^3$ \\
Disdier et al.~\cite{PhysRevLett.82.1454}       & 30&$^a$  & 300.&  5& & 400. & CD$_2$ solid &  10$^7$ \\
Izumi et al.~\cite{PhysRevE.65.036413} & 50& & 500. &  5&4 & 5.5 & C$_8$D$_8$ solid & 8.8$\times$10$^5$ \\
Belyaev et al.~\cite{belyaev} & 12& & 1500. & 15& & 200. &  CD$_2$ solid & 10$^5$ \\
Norreys et al.~\cite{norreys} & 20&$^a$  & 1300. &  15& & 120. & C$_8$D$_8$ solid & 1.2$\times$10$^8$ \\
Yamanaka et al.~\cite{PhysRevA.6.2335} & 5& &  2 10$^6$ & 100& & 100.$^b$ &D$_2$ solid &  3$\times$10$^2$ \\
& 12& &  2 10$^6$ & 100& & 100.$^b$ &D$_2$ solid &  5$\times$10$^3$ \\
& 20& &  2 10$^6$ & 100& & 100.$^b$ &D$_2$ solid &  2$\times$10$^4$ \\
Shearer et al.~\cite{shearer} & 20& &  2 10$^6$ & 80& & 100.$^b$ &CD$_2$ solid &  5$\times$10$^3$ \\
& 70& &  5 10$^6$ & 80& & 100.$^b$ &CD$_2$ solid&  10$^4$ \\
\\
Ditmire et al.~\cite{natureDitmire2} &  0&12 & 35.&  200& &  & D$_2$ cluster &  10$^4$ \\
Fritzler et al.~\cite{PhysRevLett.89.165004} &  62& & 1000. & 20& &   & D$_2$ cluster & 10$^6$ \\
Hartke et al.~\cite{hartke} & 0&2  & 40.& 200& &  & D$_2$ cluster & 2$\times$10$^3$ \\
Zweiback et al.~\cite{PhysRevLett.85.3640} & 0&12 & 35. & 200& &  & D$_2$ cluster & 5$\times$10$^3$ \\ 
\hline
\hline
\end{tabular}
}
\end{center}
\label{tab:1}
\end{table*}%
The neutron yields for the reactions $^{11}$B($p,n$)$^{11}$C and $^{63}$Cu($p,n$)$^{63}$Zn are converted 
from the radioisotope activities reported~\cite{robson,ledingham}.
Assuming a beam-target fusion mechanism, the yield $N_f$ is evaluated~\cite{kb-comnusp,PhysRevA.70.053201} as:
\begin{equation}
	\label{eq:nf}
 	N_f= \langle \sigma v\rangle  \rho_T \rho_{pl} A d_{pr} \tau	
\end{equation}
where $\langle \sigma v\rangle $  is the reaction rate per pair of particles of the reaction of interest~\cite{clayton,nacre}.
$\sigma$ and $v$ are the reaction cross section and the velocity of the colliding nuclei, respectively. 
The term $\langle \sigma v\rangle$ takes into account that the accelerated ions have an energy spread which could be 
characterized by a temperature.    
We mention that $\langle \sigma v\rangle$ is modified from the thermonuclear reaction rate, in order to take into account that
one of the colliding nuclei is at rest, as it is defined in~\ref{sec:rrppA}.; 
$\rho_T, \rho_{pl}$ and $A$ are the number densities of 
the target, plasma, and the laser spot size, respectively. $d_{pr}$ is an effective target thickness.  
For a thin target $d_{pr}$ is the thickness of the target~($d$ denotes the target thickness.) and for a thick target 
$d_{pr}$ is the projected range of the plasma ions in the target medium. 
In Eq.~(\ref{eq:nf}) the terms which can be dependent on the laser pulse energy are $\langle \sigma v\rangle $,   $\rho_{pl}$ and 
$d_{pr}$. 
If hydrodynamic flow~\cite{csernai} would be at play, we could define suitably scaled quantities which should then be scaling invariant depending 
on some dimensionless quantities such as the Reynolds number~($Re$)~\cite{csernai,PhysRevLett.59.630}.  
Classically we can define the Reynolds number as the ratio of a characteristic length $L$ and the mean free path $\lambda$. 
We can further write $L$ as the product of a typical velocity $v$~(e.g., the thermal velocity) and a typical time $\tau$~(e.g., the laser pulse duration).  
Thus 
\begin{equation}
Re=v \tau \sigma \rho_{pl}.
\end{equation}
These are the typical quantities we have to scale, in order to obtain a universal behavior.
We divide Eq.~(\ref{eq:nf}) by the terms which are clearly independent on the laser pulse energy, i.e.,  $(\tau A d \rho_T)$ and get:
\begin{equation}
	\label{eq:nf2}
 	N_f/\tau/A/d/\rho_{T}= \langle \sigma v\rangle  \rho_{pl} (d_{pr}/d)\propto Re/\tau.
\end{equation}
Eq.~(\ref{eq:nf2}) has a dimension of (Number($N$)/Time($t$)/Length($L$)$^2$/$L\times L^3$)=($N/t$)
and it is exactly the same as the scaled Eq.~(\ref{eq:nf00}) divided by a constant ($d\rho_T$). 
We, therefore, plot $\Pi$, which is defined by the l.h.s of Eq.~(\ref{eq:nf2}), in Fig.~\ref{fig:T} 
as a function of $E_L/\tau$ for the three nuclear reactions of interest. 
\begin{equation}
	\label{eq:nf3}
 	\Pi = N_f/\tau/A/d/\rho_{T}.
\end{equation}
The $\Pi$ for all reactions shows a clear power law dependence on $E_L/\tau$.
Here we used $d$=2~mm for the cluster targets~\cite{utanr}.
Given Eq.~(\ref{eq:nf2}), the power law dependence suggests that 
the reaction rate $\langle \sigma v\rangle $ for all the three reactions are similar at the temperature of interest. Especially for 
the D($d,n$)$^3$He reaction with the cluster target and with the solid target the reaction rate $\langle \sigma v\rangle $ is, in principle, identical for a given temperature. 
The power law dependence implies also that 
the plasma density~($\rho_{pl}$) is common to all cases.
In other words, the Reynolds numbers should be similar for all those reactions.  The scattering of data observed in the plot might be due to some differences in the experimental set-ups not considered here, e.g., in the plot, we have assumed that $d_{pr}$ is larger than the (secondary) target thickness $d$, 
so that we have replaced $d_{pr}$ by $d$. 
Repeating the measurement under common experimental conditions,  for instance using always very thin targets, might reduce somewhat the scattering of the data.
In such a way one could easily scale the relevant features.  
Nevertheless the plot of Fig.~\ref{fig:T} indicates scaling over almost as many as 12 orders of magnitude. 
Deserving special mention in Fig.~\ref{fig:T} is the density of the cluster gas target
which we set to 5 $\times$ 10$^{18}$ cm$^{-3}$, as given in~\cite{td88}.
However, the local density within a cluster is higher than it and it is close to the solid density ~(10$^{23}$ cm$^{-3}$)~\cite{zweiback}.
Fig.~\ref{fig:Cr} shows the $\Pi$ divided by the plasma critical density $\rho_{cr}=10^{21}$ cm$^{-3}$.
In the figure  the cluster target data are divided by an additional factor 8,
to get a better coincidence between the results of the cluster target and the solid target.
This factor could be attributed either to an ambiguity of target thickness, which we used 2~mm as mentioned above or  
to an ambiguity of the number density of the cluster target, or to both of them. 
Notice that for a real gas target the number of fusions would be very small since it is transparent to the laser pulses.
In Fig.~\ref{fig:Cr} the curve is the result of the best fit to all the data for the reaction D($d,n$)$^3$He in Tab.~\ref{tab:1} and 
from Ref.~\cite{utanr}. The curve has
a power law dependence with an exponent 2.3 and it is
\begin{eqnarray}
 \Pi/\rho_{cr}&=& \langle \sigma v\rangle  (\rho_{pl}/\rho_{cr}) (d_{pr}/d)\nonumber \\
 &\sim& 2.1 \times 10^{-14} (E_L/\tau)^{2.3} (cm^3 s^{-1}),
 \label{eq:parDdn} 
\end{eqnarray}
where $E_L/\tau$ is in units of 10$^{15}$ W and it is so throughout this paper unless otherwise specified. 
For the reaction $^{63}$Cu($p,n$)$^{63}$Zn the best fit is given by 
\begin{equation}
 \Pi/\rho_{cr}\sim 7.9 \times 10^{-14} (E_L/\tau)^{2.3} (cm^3 s^{-1}).
 \label{eq:parCupn} 
\end{equation}
with the same exponent 2.3 as the reaction D($d,n$)$^3$He.
A careful observation of Fig.~\ref{fig:Cr} reveals
that the slope for the reaction $^{11}$B($p,n$)$^{11}$C is steeper than 
the slope for the reaction D($d,n$)$^3$He.   The best fit for the data of the reaction $^{11}$B($p,n$)$^{11}$C 
gives a power law dependence with an exponent 3.5:
\begin{equation}
 \Pi/\rho_{cr} \sim1.3 \times 10^{-10} (E_L/\tau)^{3.5} (cm^3 s^{-1}). 
  \label{eq:par11Bpn}
\end{equation}
The difference in the power law dependences could be attributed to the difference in the reaction Q-values, as we will see in the following discussions.

\section{Origin of the scaling law in the laser driven fusion yield}
\label{sec:osl}
To investigate the origin of the slope of the scaling in the $\Pi/\rho_{cr}$, 
we reformulate the scaling laws for the maximum energy of laser-accelerated protons from thin foil targets
and in the conversion efficiency of the laser pulse energy to the proton beam energy. 
Eq.~(\ref{eq:kTElap1}) derived in~\ref{sec:mpe_pt} gives the relation between the plasma temperature and the laser pulse energy:
\begin{equation}
\label{eq:kTEl}
kT\sim \sqrt{\frac{E_L}{\tau}} \tanh^2 \left( \frac{c\tau}{R}\sqrt{\frac{m_e}{m_p}} \left(\frac{\eta}{8.7 \times 10^{-6}} \frac{E_L}{\tau} \right)^{0.25}\right),
\end{equation}
where $m_e$, $m_p$,$c$ and $\eta$ are the electron andproton masses, the speed of light and the conversion efficiency of the laser pulse into the hot electrons, respectively, as they are determined in~\ref{sec:mpe_pt};  we have assumed that the plasma temperature is proportional to the maximum proton energy.
The constant factor which should be multiplied to this equation to obtain the plasma temperature could be determined by using experimental data
of the energy spectra of accelerated protons. The spectra are measured by using either radiochromic film~(RCF) stacks or a CR-39 detector or a Faraday cup with Thomson 
parabola,  which are placed behind the primary target.  
In Tab.~\ref{tab:0} we summarize the laser parameters and the temperature of accelerated protons 
available from metal foil target irradiations at the VULCAN laser,
at Calisto and Titan lasers at LLNL, at ASTRA Ti:Sapphire laser at the RAL
and the laser facility at the Korea Atomic Energy Research Institute~(KAERI).
\begin{table*}[htdp]
\caption{The number and the temperature of accelerated protons in ultrahigh-intensity laser-matter interactions at selected laser facilities worldwide. 
The number and the temperature of accelerated ions are from the references if those are given explicitly. Otherwise  those are obtained by fitting 
the proton spectra given in the references and with an assumption that protons are emitted in a cone of angle 22 deg.~[4].  $^a$coated with ErH$_3$ on the back.}

\begin{center}
{\footnotesize
\begin{tabular}{cc|rrcrc}
\hline\hline
    & & $E_L$(J) &  $\tau$(fs)  & Target (thickness) & \multicolumn{1}{c}{$N_i$} & $kT$ (MeV)\\     
\hline
Calisto at LLNL & \cite{foord}&     10 &      100  &  Au (15~$\mu$m) &        2$\times$10$^{11}$ & 1.4  \\
Titan at LLNL & \cite{hey} &     150 &      600  &   Au (14~$\mu$m)$^a$ & 1.2$\times$10$^{13}$ &  3.3 \\
VULCAN at the RAL & \cite{ledingham} &  300  &      750  & Al (10 $\mu$m)  &     1$\times$10$^{12}$  & 5.0 \\  
                                      &  \cite{spencer2} &  120  &      1000  &  Al (unknown)   &   1$\times$10$^{12}$  & 2.9 \\  
ASTRA at the RAL & \cite{spencer} &   0.2  &     60  &   Mylar (13 $\mu$m)   &  4.$\times$10$^{11}$  & 0.03\\  
                                  &  &  0.2  &     60  &   Mylar (23 $\mu$m)   &  2.$\times$10$^{10}$  & 0.13\\  
laser at the KAERI  & \cite{lee} &   0.3  &     30  &    Al (15 $\mu$m)   &  1.9$\times$10$^{11}$  & 0.03\\  
                                         &                  &   0.3  &     30  &    Mylar (13 $\mu$m)   &  1.5$\times$10$^{11}$  & 0.06\\  
\hline
\hline
\end{tabular}
}
\end{center}

\label{tab:0}
\end{table*}%
\begin{figure}[]
  \centering
  \includegraphics[height=.38\textheight, bb=0 0 846 594]{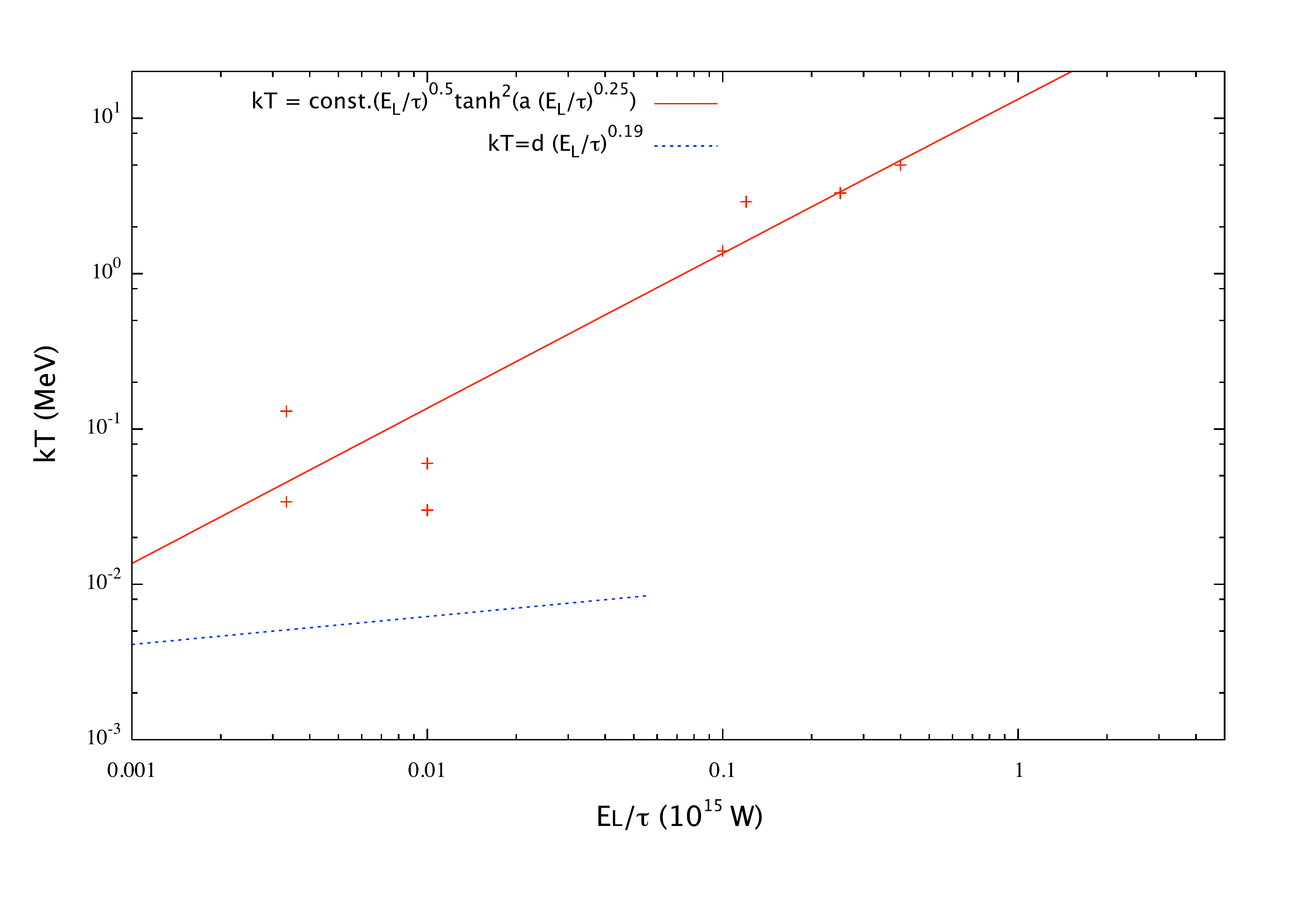}

  \caption{Temperature of the accelerated ions from the laser-solid interaction 
  as a function of $E_L/\tau$. The experimental data summarized in Tab.~2 (plusses) are shown with
  a fit to the data with Eq.~(A.1) (solid curve).
  A power law dependence of the accelerated ion temperature from the laser-cluster interaction~[24]~(dashed curve).  
  }
  \label{fig:KTSc}
\end{figure}
In Fig.~\ref{fig:KTSc} we plot the temperature of the plasma in the table 
as a function of $E_L/\tau$. 
The fitting curve shown in Fig.~\ref{fig:KTSc} by the solid line gives
the coefficient of Eq.~(\ref{eq:kTEl}) to be 
3.4 $\times$ 10$^2$, i.e., 
the temperature of the plasma is thus related to the laser power ($E_L/\tau$) by 
\begin{equation}
\label{eq:kt2}
kT= 3.4\times 10^2 \sqrt{\frac{E_L}{\tau}} \tanh^2 \left( \frac{c\tau}{R}\sqrt{\frac{m_e}{m_p}} \left(\frac{\eta}{8.7 \times 10^{-6}} \frac{E_L}{\tau} \right)^{0.25}\right).
\end{equation}
Although systematical measurements are required for a more precise determination of this coefficient. 
Eq.~(\ref{eq:kt2}) is simplified in two limits, where either the pulse duration ($\tau$) is much shorter/longer than the time in which a proton remains in the vicinity of the accelerating surface charge ($\tau_0$), as 
\begin{eqnarray}
kT &\sim& \eta (E_L/\tau)  \ (for \  \tau \ll 2\tau_0) \label{eq:linear} \\
 &\sim& \sqrt{\eta E_L/\tau}  \ (for \ \tau \gg 2\tau_0) \label{eq:rootsquare}.
\end{eqnarray}
In other words, the plasma temperature of laser-accelerated protons from thin foil targets 
depends on the laser power linearly in the limit ($\tau \ll 2\tau_0$) 
or the plasma temperature follows square-root scaling of the laser power in the other limit  ($\tau \gg 2\tau_0$). 
We note that the lasers with the pulse power higher than 100~TW correspond to the latter limit and that 
the majority of the experimental data in Fig.~\ref{fig:Cr} are in this laser-power region.   Given this condition,
and that the original equation (\ref{eq:kt2}) is difficult to invert,
we use Eq.~(\ref{eq:rootsquare}) to express the relation between the laser power and the plasma temperature.
Fitting of the data gives the coefficient of Eq.~(\ref{eq:rootsquare}):
\begin{equation}
\label{eq:rootsquare7}
kT=7.0 \sqrt{E_L/\tau}.
\end{equation}
The coefficient of this equation has a unit of (MeV/W$^{0.5}$).

In the case of laser-irradiation on deuterated cluster target, 
the average energies of emitted deuterons from D$_2$ and CD$_4$ cluster plasmas are known to 
scale as~(Fig. 4 in \cite{td74}) 
\begin{equation}
 \bar{E}_{ion} \sim const. \times E_L^{0.19}  (keV).
\end{equation}
In the experiments the pulse length is fixed at 100 fs 
and the emitted ion spectra have been measured by using two Faraday cups, placed about 52~cm and 36~cm from the laser 
focus and at different angles.
The temperature of the deuteron plasma is related to the ions average energy by $kT=\frac{2}{3}\bar{E}_{ion}$. 
This leads to the power law dependence of the temperature of deuteron plasma on the laser power:
\begin{equation}
\label{eq:kt3}
kT\sim 1.5 \times 10^{-2} \times (E_L/\tau)^{0.19} (MeV).  
\end{equation}
The curve of the power law dependence is shown by the dashed curve in Fig.~\ref{fig:KTSc}.
Since the plasma is most probably out of equilibrium and rapidly expanding, the temperature 
measured from the plasma ion distributions could be different from 
the temperature at the time when nuclear fusions occur, in the case of deuterated cluster irradiation.
Nevertheless the comparison between the two curves
in Fig.~\ref{fig:KTSc} shows that 
the power law dependence of the plasma temperature in deuterated cluster irradiation is clearly 
weaker than that of the proton plasma temperature in the solid target irradiation.  

\begin{figure}[]
  \centering
  \includegraphics[height=.38\textheight, bb=0 0 846 594]{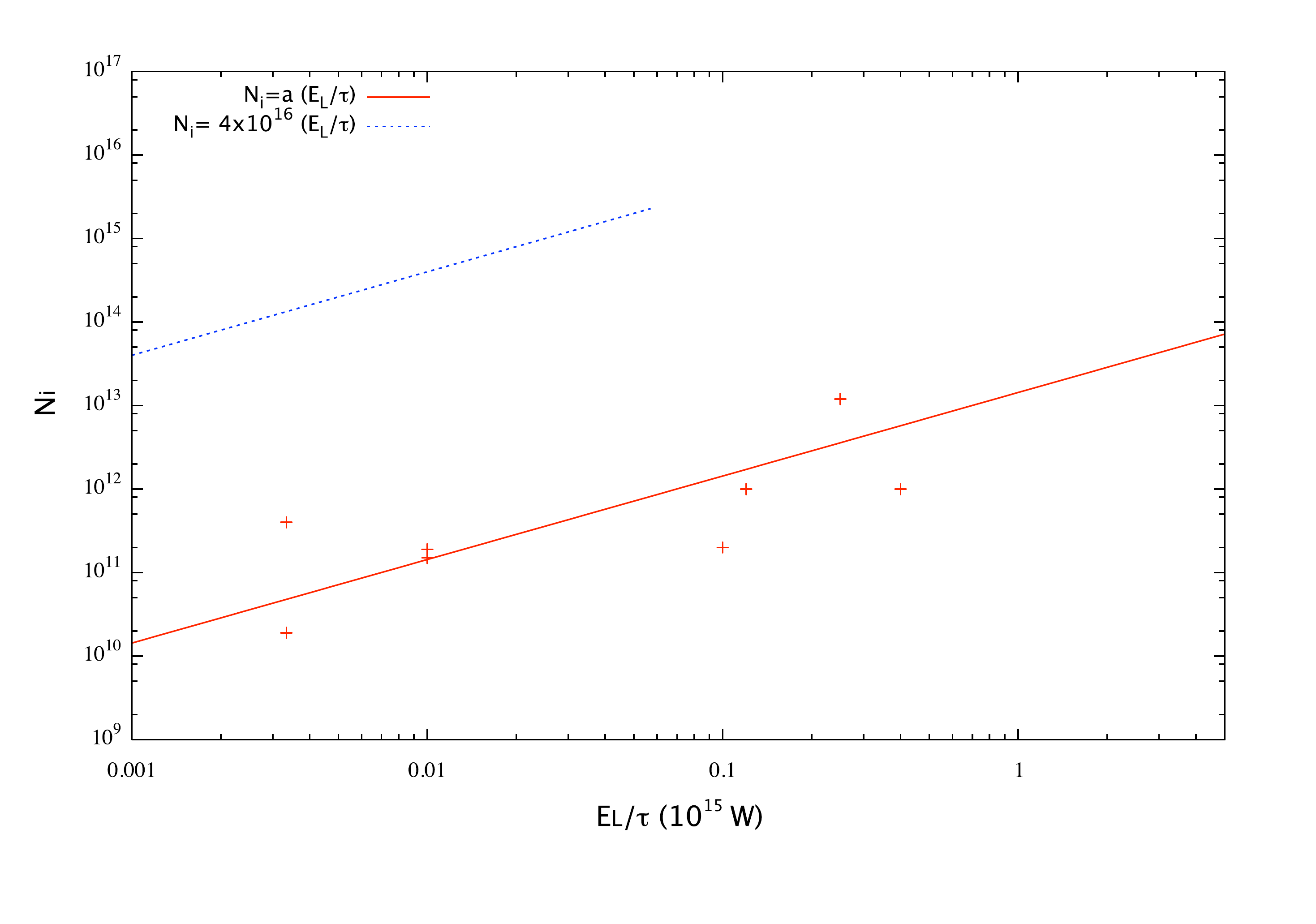}
    
  \caption{Number of accelerated ions from the laser-solid interaction as a function of $E_L/\tau$.
   Experimental data at the laser facilities listed in Tab.~2 (plusses).
   A fitting~(solid curve) to the data is shown and it is $N_i=1.4 \times 10^{13} E_L/\tau$.
  The power law dependence on the laser power of the number of deuterons emitted from D$_2$ cluster~[24] (dashed curve). 
  }
  \label{fig:NiaSc}
\end{figure}

Meanwhile the conversion efficiency of the laser pulse energy to the proton beam energy 
depends on the laser pulse energy linearly
(Fig. 2 in~\cite{robson}). There
the conversion efficiency is determined by integrating the energies of all the protons accelerated above~4 MeV and 
it is, therefore, proportional to the number of accelerated protons~($N_i$): 
\begin{equation}
 	\label{eq:ni}
  N_i \sim (E_L/\tau).
\end{equation}
To verify our assumption on the $E_L/\tau$ dependence of the number of accelerated ions for different pulse-length irradiations, 
in Fig.~\ref{fig:NiaSc} we plot the data summarized in Tab.~\ref{tab:0} as a function of $E_L/\tau$. 
It is well known that the number of accelerated ions depends not only on the laser parameters but also 
on the density, the thickness and the surface treatment of the target irradiated by lasers~\cite{spencer,lee,hey}. 
The scattering of data points in the figure could be reduced by fixing the experimental conditions of the targets.
The fitting curve to the data of the number of the laser-accelerated protons from thin foil targets 
is shown by the solid curve 
in Fig.~\ref{fig:NiaSc}:
\begin{equation}
N_i=1.4 \times 10^{13} E_L/\tau.
\end{equation}
The coefficient of this equation has a unit of (W$^{-1}$). 

In the case of deuterated-cluster target irradiation, 
the number of accelerated deuterons is, indeed, known to
scale as a function of the laser pulse energy for a fixed pulse duration~\cite{PhysRevA.70.053201}, i.e.,
$N_i=$4.3$\times$10$^{14}E_L^{1.1}$ 
and $N_i=$3.4$\times$10$^{14}E_L^{0.94}$, where the deuteron ion yield is determined using the two Faraday cups which give 
two power law dependencies.  
Because of the fixed pulse length ($\tau=100$~fs) in the measurement, $N_i$ is proportional to the $E_L/\tau$, as well.
As a function of the laser power it is given by 
\begin{equation}
 \label{eq:nicl}
   N_i \sim 4. \times 10^{16} E_L/\tau.  
\end{equation}
In Fig.~\ref{fig:NiaSc} this equation is shown by the dashed curve.

At last, 
the density of the plasma is, approximately, proportional to the number 
of the accelerated ions, i.e., 
\begin{equation}
\label{eq:rhopl}
\rho_{pl} \sim N_i \sim E_L/\tau.
\end{equation}

Now we discuss the two terms $\langle \sigma v\rangle$ and $(d_{pr}/d)$. 
An important factor that should be taken into account to test the scaling laws 
is the Q-value of the reactions. In 
the case of the three reactions of interest, 
the latter two reactions have negative Q-values, while the first one has a positive Q-value. 
This might change slightly the slopes of $\langle \sigma v\rangle$, since the incident energy of the reaction 
must overcome the threshold energy,
or alternatively the low energy ions give no contribution to the yields because of the Q-value.  
We begin with the reaction rate at high temperatures,
as in the case of the reaction $^{63}$Cu($p,n$)$^{63}$Zn observed at the VULCAN laser facility.
The reaction cross section above the Coulomb barrier is approximated~\cite{isosp,takigawa} by
 \begin{equation}
 	\sigma_R=\pi R^2 \sqrt{1-\frac{V_c}{E}},
\end{equation}
where $R, V_c$ and $E$ are the sum of the nuclear radii of the colliding nuclei, the height of the Coulomb barrier 
between the colliding nuclei and the incident energy of the collision. 
Furthermore we approximate the incident energy of collision by the maximum proton energy $E_{pmax}$, which
follows the same scaling relation as the plasma temperature, i.e.,
with an intermediate value between linear and square-root of the laser pulse energy.
 (Eq.s~(\ref{eq:linear}) and~(\ref{eq:rootsquare})).
$\langle \sigma v \rangle$ is, then, approximated by 
 \begin{eqnarray}
 	\sigma _R \sqrt{\frac{2E_{pmax}}{m_p}}&\sim& \pi R^2 \sqrt{1-\frac{V_c}{E}} \sqrt{\frac{2 \times (E_L/\tau)^{0.75\pm 0.25}}{m}} \nonumber  \\
	&\sim&  (E_L/\tau)^{0.38\pm 0.13} \label{eq:sv}
 \end{eqnarray}
that is the leading term has a power law dependence with an exponent $0.38\pm 0.13$.  
The curve for the proton projected range in a copper foil is given as a function of the incident energy $E$ 
by the SRIM code~\cite{srim}. In the energy region $E > 1$~MeV, it is approximated by 
\begin{equation}
    d_{pr}\sim E^{1.74}.
\end{equation}
Here again we replace the incident energy $E$ by the maximum proton energy.
We, thus, obtain a relation between laser pulse energy and the projected range:
\begin{equation}
	\label{eq:dpr}
    d_{pr}\sim \left((E_L/\tau)^{0.75\pm0.25}\right)^{1.74} \sim (E_L/\tau)^{1.31\pm0.44}.
\end{equation}
Finally, by using Eq.s~(\ref{eq:rhopl}), ~(\ref{eq:sv}) and~(\ref{eq:dpr}), 
the $\Pi/\rho_{cr}$
for the reaction $^{63}$Cu($p,n$)$^{63}$Zn has the $E_L/\tau$-dependence:
\begin{eqnarray}
 	\Pi/\rho_{cr}&=& \langle \sigma v \rangle (\rho_{pl}/\rho_{cr}) (d_{pr} /d) \nonumber  \\
	&\sim& (E_L/\tau)^{0.38\pm 0.13}  (E_L/\tau)  (E_L/\tau)^{1.31\pm0.44} \times const. \nonumber \\
	&\sim& (E_L/\tau)^{2.69\pm0.57}.
\end{eqnarray}
The obtained $(E_L/\tau)$-dependence from this rough estimate is 
in accord with the power $(E_L/\tau)^{2.3}$, found in Eq.~(\ref{eq:parCupn}).
Furthermore it is clearly seen that the exponent 2.3 is closer to a limit (2.69-0.57=2.12), which corresponds to the 
root-square scaling of the plasma temperature on the laser pulse power, rather than the other limit (2.69+0.57=3.26).
This fact supports our approximation of Eq.~(\ref{eq:kt2}) by Eq.~(\ref{eq:rootsquare}).

\bigskip
\section{Formalism}
\label{sec:rrpp}
The rough estimate given at the end of the preceding section shows an excellent agreement with the observed power law dependence. 
We derive the scaling of the fusion yield more generally and precisely in this section. 

\subsection{Projected range and reaction rate}
\begin{figure}
  \includegraphics[height=.37\textheight, bb=0  0 846 594]{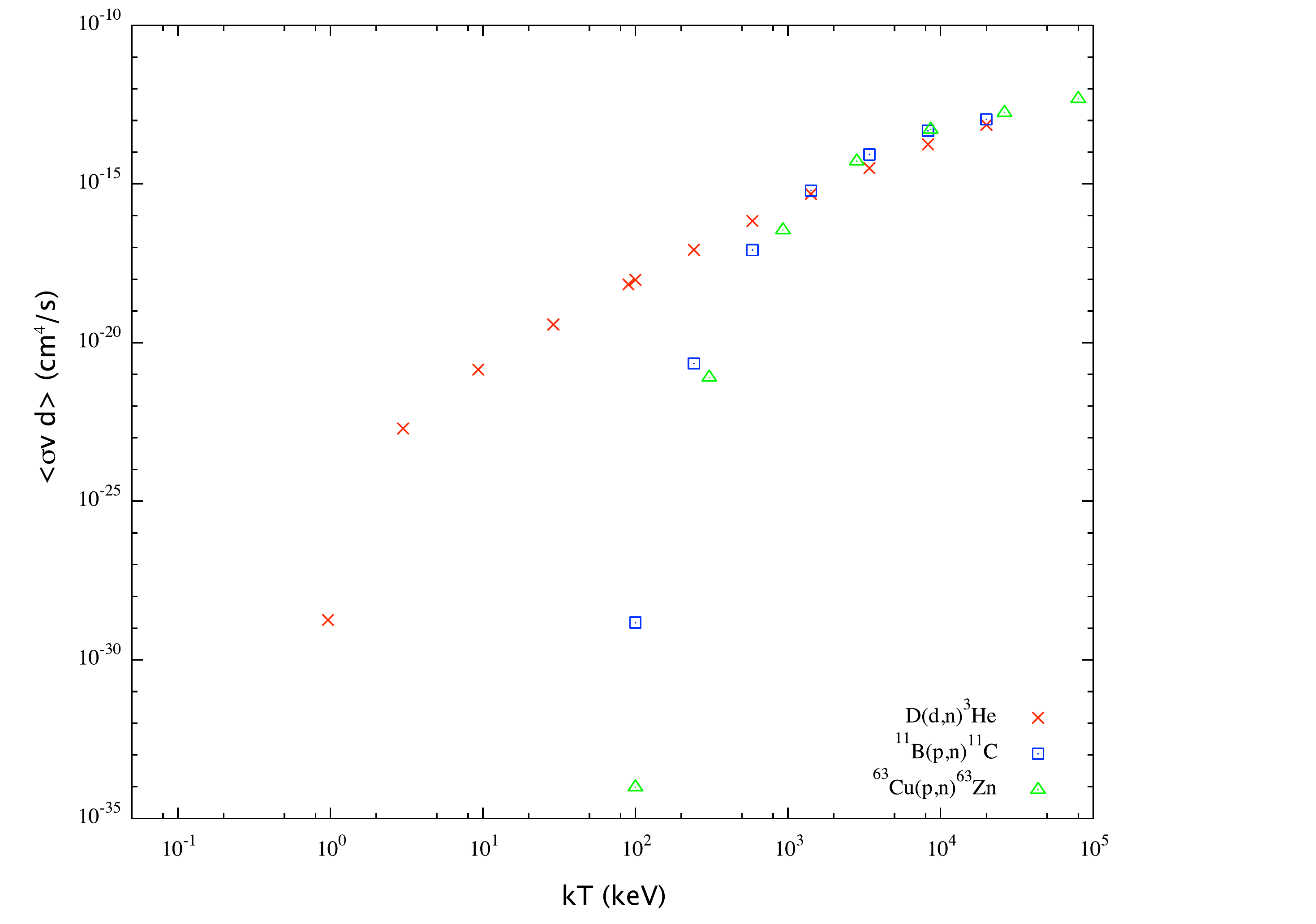}
   \includegraphics[height=.37\textheight, bb=0  0 846 594]{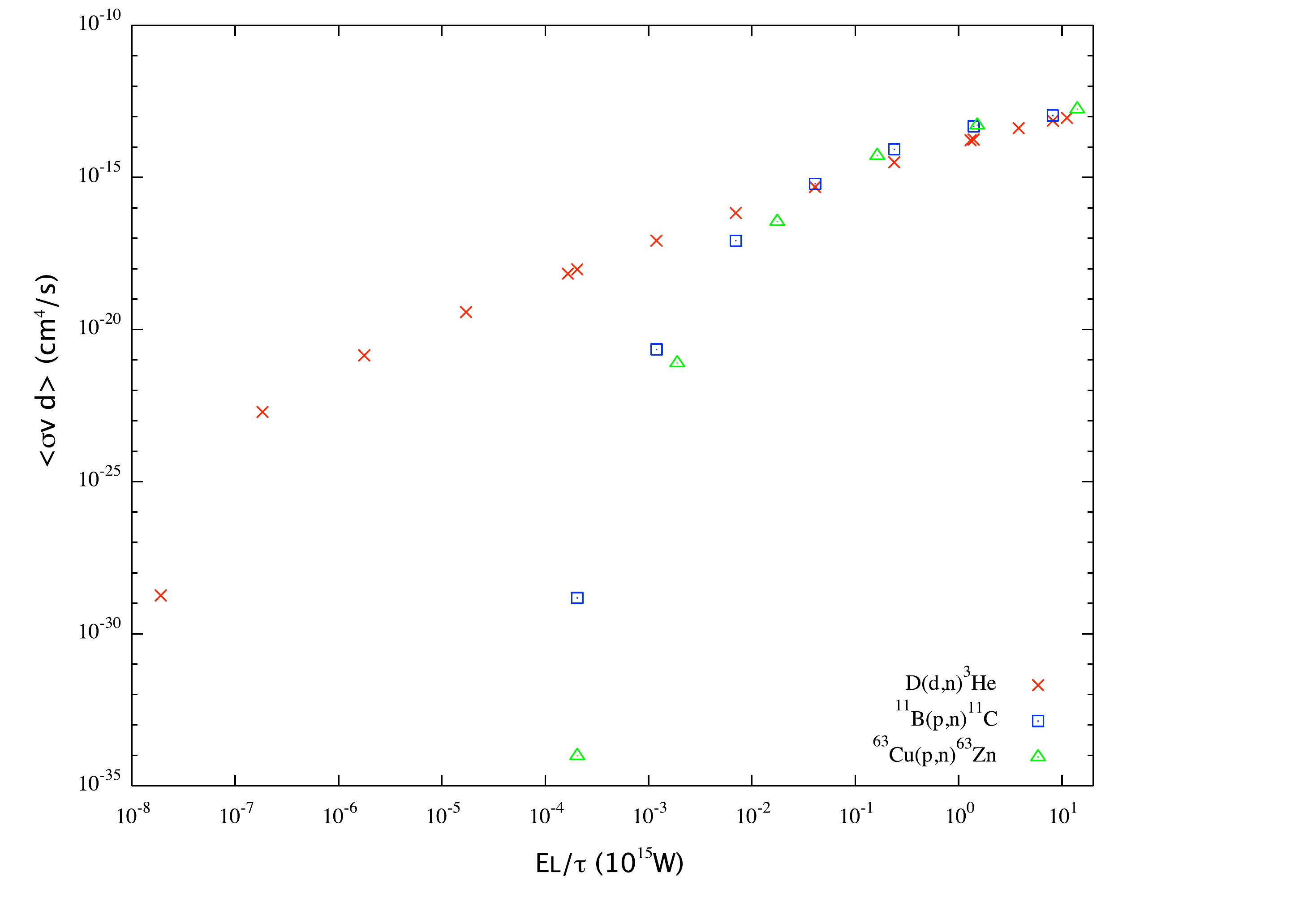}
  \caption{ \label{fig:SVD} (Top panel) The $\langle \sigma v d \rangle$ defined by Eq.~(27) as a function of the plasma temperature. 
  For the reactions 
  D($d,n$)$^3$He~(crosses), $^{11}$B($p,n$)$^{11}$C~(squares) and $^{63}$Cu($p,n$)$^{63}$Zn~(triangles). 
  The cross sections are retrieved from the NACRE~[44], EXFOR~[55] databases. The stopping power data are
  obtained using the SRIM code~[51]. (Bottom panel) The same with the top panel but as a function of the laser power, 
  derived with Eq.~(14).  }
\end{figure}

In Sec.~\ref{sec:sfy}, the yield is expressed in terms of the projected range~($d_{pr}$).
That is the cross section is assumed to be a constant at the given incident energy, through the whole target. 
This assumption is correct, if the target is thin and the energy loss of the accelerated ions in the target is negligible. 
While for a thick target the yield at an incident energy ($E_{in}$) is given by integrating the beam energy down to zero and it is
proportional to~\cite{stave,angulo}  
\begin{equation}
\label{eq:csstp}
\int_{E_{in}}^0 \frac{\sigma(E')}{S_p(E')} dE', 
\end{equation} 
where $S_p(E')$ is the stopping power of accelerated ions in the (secondary) target.  
This quantity takes into account 
the projectile energy loss in the target. 
We evaluate the reaction rate for the plasma ions colliding with ions at rest which is defined by Eq.~(\ref{eq:rrE}) in~\ref{sec:rrppA},
but with the substitution of $\sigma(E)$ by Eq.~(\ref{eq:csstp}):
\begin{equation}
	\langle \sigma v d \rangle = \int dE_{in}  \left( \int_{E_{in}}^0 \frac{\sigma(E')}{S_p(E')} dE' \right) \sqrt{\frac{2E_{in}}{m_1}} \psi(E_{in}), \label{eq:SVD}
\end{equation}
where $ \psi(E_{in})$ is the energy spectrum of the plasma ions.
This gives the definition of $\langle \sigma v d \rangle$, which is a function of the temperature $kT$.
In terms of $\langle \sigma v d \rangle$, $\Pi/\rho_{cr}$ is rewritten as 
\begin{equation}
\label{eq:svd}
\Pi/\rho_{cr}=\langle \sigma v \rangle (\rho_{pl}/\rho_{cr})(d_{pr}/d)=\langle \sigma v d \rangle (\rho_{pl}/\rho_{cr})d^{-1}.
\end{equation} 
The top panel in Fig.~\ref{fig:SVD} shows $\langle \sigma v d \rangle$ for the three reactions as a function of the plasma temperature $kT$. 
As derived in the preceding section, the plasma temperature of the laser-accelerated protons from thin foil targets
is related to the laser pulse power by a square-root dependency (Eq.~(\ref{eq:rootsquare7})).
By using this equation we obtain the ($E_L/\tau$)-dependence of the $\langle \sigma v d \rangle$.
In the bottom panel of Fig.~\ref{fig:SVD} 
the ($E_L/\tau$)-dependence of the $\langle \sigma v d \rangle$ is shown
for the three reactions induced by laser-accelerated protons from thin foil targets. 
Specially for the case of the D$_2$ and CD$_4$ cluster gas targets, the relation between the plasma temperature 
and the laser power is different from Eq.~(\ref{eq:rootsquare7}).
For the cluster gas targets Eq.~(\ref{eq:kt3}) can be used to obtain the ($E_L/\tau$)-dependence of the $\langle \sigma v d \rangle$.

\subsection{Plasma density}
\label{sec:plasmadens}
The density of the plasma can be dependent on the temperature of the plasma.
To our knowledge, there is no available experimental data on the plasma
temperature~(the laser pulse energy) dependence of the plasma density.
Here we derive the laser pulse energy dependence of the plasma density
in two limits, using the power law dependence of the number of accelerated ions~(Eq.~(\ref{eq:ni})). 
The first limit is the case where the volume of the plasma ($V_{pl}$) has no $E_L/\tau$-dependence. 
The plasma density is related to the laser pulse power linearly:
\begin{equation}
\rho_{pl}  = \frac{N_i}{V_{pl}} \sim E_L/\tau.
\label{eq:rpl1}
\end{equation}
The other limit is a case where the volume of the generated plasma has the laser-pulse power dependence, 
e.g., $V_{pl}$ is expressed as a product of 
the laser focal spot size and 
the range covered by the accelerated ions in a period of the laser pulse duration, i.e., $V_{pl}=A \tilde{d}.$
This range can be approximated by $\tilde{d} \sim v_{max} \tau = \sqrt{2E_{pmax}/m_p} \tau.$ Given the 
laser-pulse power dependence of the maximum proton energy $E_{pmax} \sim  (E_L/\tau)^{0.5}$~(Eq.(\ref{eq:sv})),
the volume of the plasma is approximated by $V_{pl}=A \tau (E_L/\tau)^{0.25}$. 
The approximation is justified because the energy loss of the energetic ions in the plasma is known to be negligible~\cite{prl2011}. 
Substituting this, the plasma density is expected to depend on the laser pulse power:
\begin{equation}
\rho_{pl}  = \frac{N_i}{V_{pl}} \sim \frac{(E_L/\tau)}{A \tau (E_L/\tau)^{0.25}} . 
\label{eq:rpl2}
\end{equation}
Which one of the two limits is closer to the real case
might be inferred from the scaled $\Pi/\rho_{cr}$.  
The scaled $\Pi/\rho_{cr}$ in Eq.s ~(\ref{eq:parDdn}),~(\ref{eq:parCupn}) and (\ref{eq:par11Bpn}) is rewritten 
in terms of $\langle \sigma v d \rangle$ as 
\begin{eqnarray}
\Pi/\rho_{cr}=\langle \sigma v d \rangle (\rho_{pl}/\rho_{cr}) d^{-1}. \label{eq:elt}  
\end{eqnarray}
In the first limit~(Eq.~(\ref{eq:rpl1})), this equation becomes 
\begin{eqnarray}
\Pi/\rho_{cr}=\langle \sigma v d \rangle (E_L/\tau)(\rho_{cr} d)^{-1}, \label{eq:elt1}  
\end{eqnarray}
where the last term of the right-hand-side is a constant. The $E_L/\tau$-dependence comes from the first two terms.
In the second limit,~(Eq.~(\ref{eq:rpl2})), 
Eq.~(\ref{eq:elt}) gives:
\begin{eqnarray}
\Pi/\rho_{cr}=\langle \sigma v d \rangle (E_L/\tau)^{0.75}(\rho_{cr} A d \tau)^{-1}.\label{eq:elt2}  
\end{eqnarray}
We evaluate Eq.~(\ref{eq:elt1}) and Eq.~(\ref{eq:elt2}) numerically. 
It should be emphasized that the laser power dependence of Eq.~(\ref{eq:elt1}) and Eq.~(\ref{eq:elt2}) are
derived using Eq.~(\ref{eq:rootsquare}).

\subsection{Discussions}
\begin{figure}
    \includegraphics[height=.38\textheight, bb=0  0 846 594]{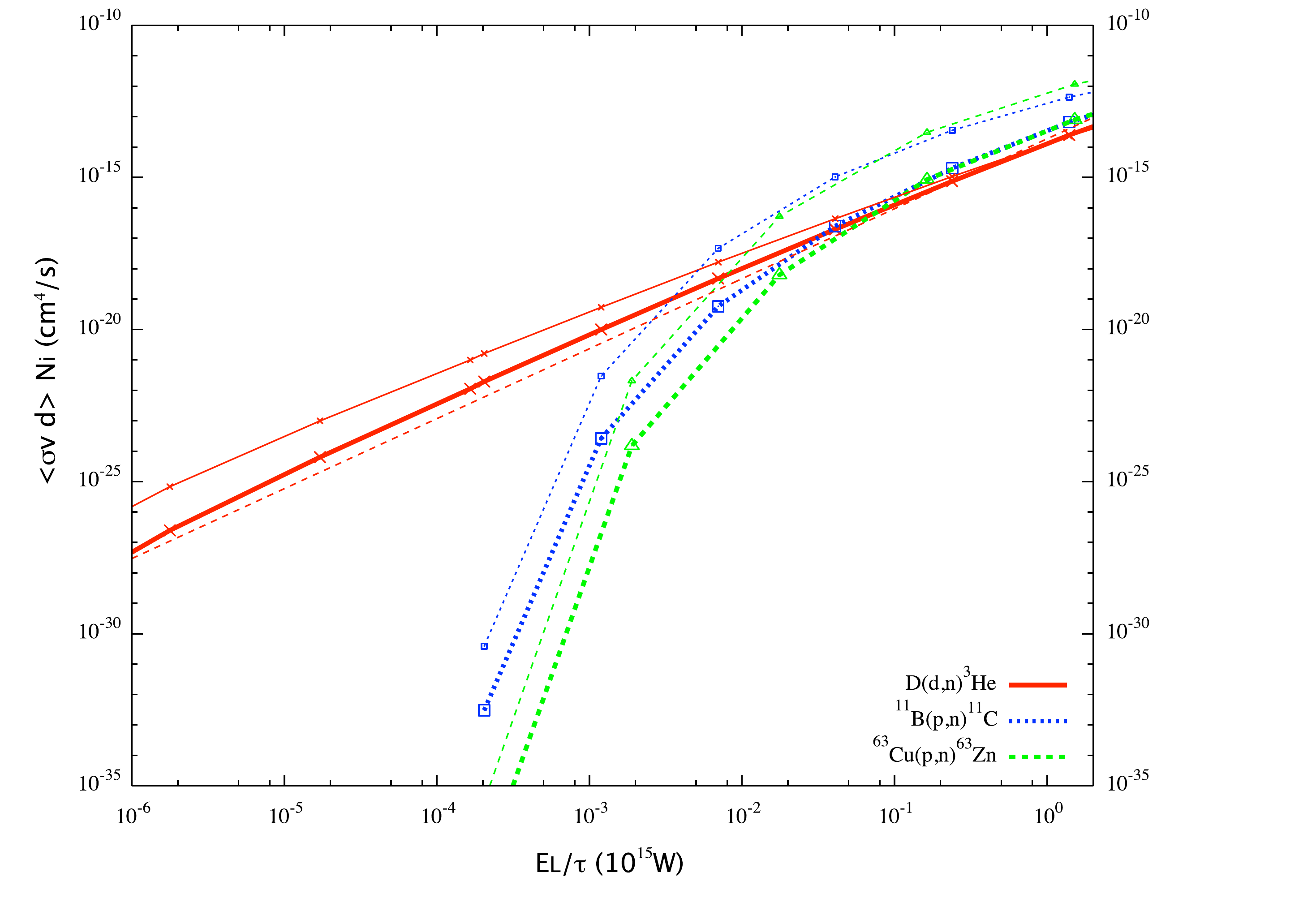}
    
    \includegraphics[height=.37\textheight, bb=0  0 846 594]{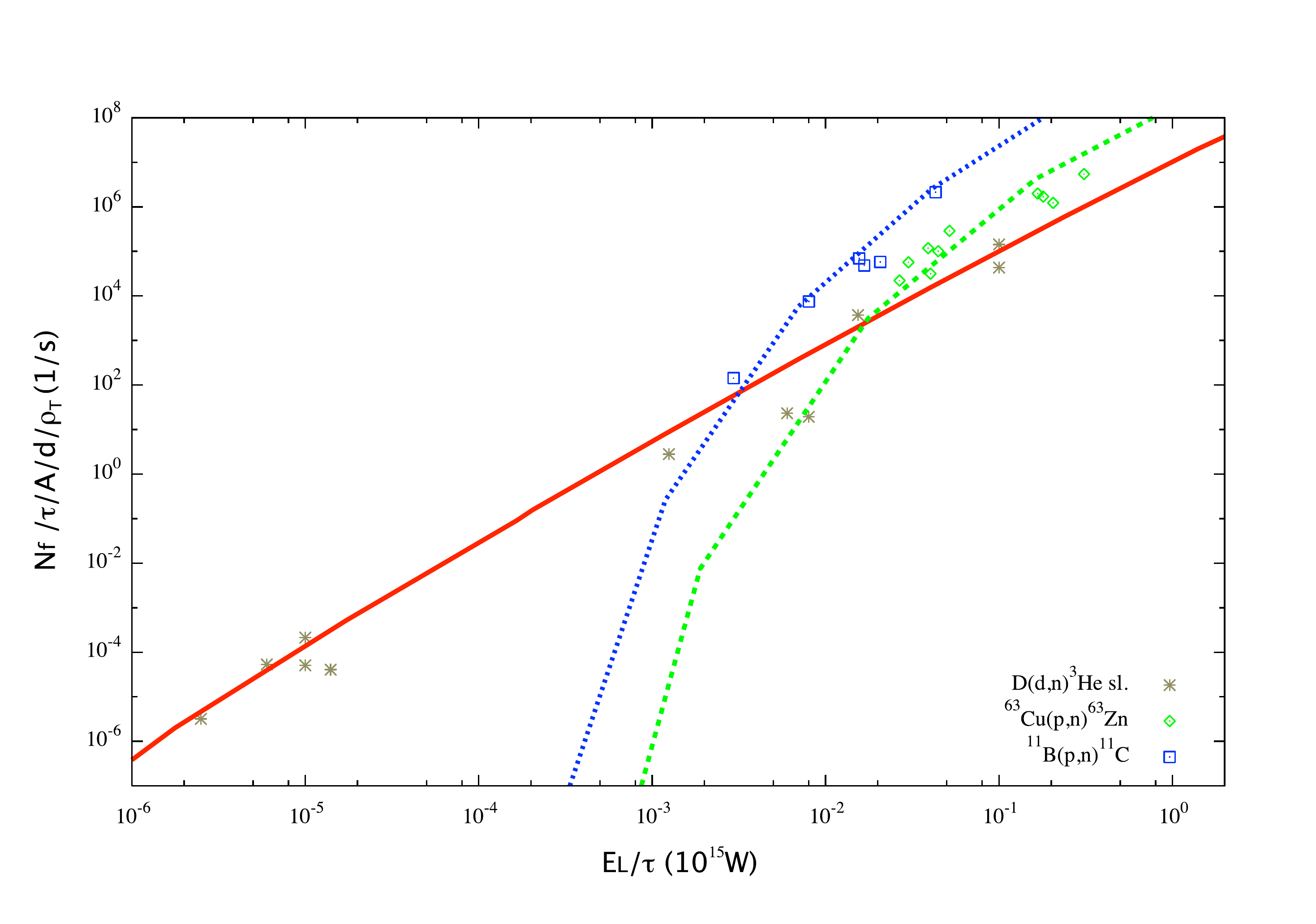}
      \caption{ \label{fig:RRdd3} (Top panel)~The laser power dependence of $\Pi/\rho_{cr}$ evaluated in two limits for the reactions D($d,n$)$^3$He~(crosses), $^{11}$B($p,n$)$^{11}$C~(squares) and $^{63}$Cu($p,n$)$^{63}$Zn~(triangles).   One limit is calculated by Eq.~(32) (bigger symbols) and the other limit is given by Eq.~(34) (smaller symbols).
  The curve of  $(E_L/\tau)^{2.3}$ is given by the straight dashed line.  The abscissa is limited in the same region of Fig.~3. 
 (Bottom panel)~The thick curves in the top panel are compared with the experimental data of the three reactions
  induced by laser-solid target interaction.  $\Pi$ of the experimental data is shown by points as they are defined in Fig.~2. }  
\end{figure}

In the top panel of Fig.~\ref{fig:RRdd3}
the $E_L/\tau$-dependent term of Eq.~(\ref{eq:elt1}) is shown by bigger symbols connected with thick lines for the reactions  
D($d,n$)$^3$He~(crosses), $^{11}$B($p,n$)$^{11}$C~(squares) and $^{63}$Cu($p,n$)$^{63}$Zn~(triangles).
The $E_L/\tau$-dependent term of Eq.~(\ref{eq:elt2}) is shown by smaller symbols connected with thin lines for the same reactions.  
The thin straight dashed line represents the curve of  $(E_L/\tau)^{2.3}$, i.e., the observed scaling relation of experimental data for the reaction D($d,n$)$^3$He
(Eq.~(\ref{eq:parDdn})).
For the reaction D($d,n$)$^3$He,
it is clearly seen that the curve with the exponent 2.3 is close to the estimate by Eq.~(\ref{eq:elt1}), 
while the estimate by Eq.~(\ref{eq:elt2}) gives an exponent less than 2.  
The comparison with the curve with the exponent 2.3 suggests that the plasma density depends linearly on the laser power, i.e., 
the volume of the generated plasma has no $E_L/\tau$-dependence. 
The same applies for the reaction $^{63}$Cu($p,n$)$^{63}$Zn in the laser power region 
3 $\times$ 10$^{13}$~W $ < E_L/\tau <$ 3 $\times$ 10$^{14}$~W, where one finds the experimental data.
While for the reaction $^{11}$B($p,n$)$^{11}$C, in the laser power region 
3 $\times$ 10$^{12}$~W $ < E_L/\tau < 5 \times$ 10$^{13}$~W, the curve given by Eq.~(\ref{eq:elt1})
has the $(E_L/\tau)$-dependence of $(E_L/\tau)^{3.6}$. This is 
in good agreement with the laser power dependence of the exponent 3.5 found in Eq.~(\ref{eq:par11Bpn}).
In the bottom panel of Fig.~\ref{fig:RRdd3} the thick curves given by Eq.~(\ref{eq:elt1}) for the three reactions
are compared with the experimental data of $\Pi$ in Fig.~\ref{fig:T}. 
In the figure Eq.~(\ref{eq:elt1}) is multiplied by a constant to be compared with each of their experimental data.
The laser power dependences of $\langle \sigma v d \rangle N_i$ derived under an assumption of 
the linear dependence of the plasma density on the laser power 
(Eq.~(\ref{eq:elt1}))
for the three reactions 
are in accord with $\Pi$ of the experimental data. 
The steep rises of the curves given by Eq.~(\ref{eq:elt1}) for the reactions $^{11}$B($p,n$)$^{11}$C and $^{63}$Cu($p,n$)$^{63}$Zn
in the laser power region less than 
5 $\times$ 10$^{13}$~W are due to the negative Q-values of the reactions.

\begin{figure}
    \includegraphics[height=.37\textheight, bb=0  0 846 594]{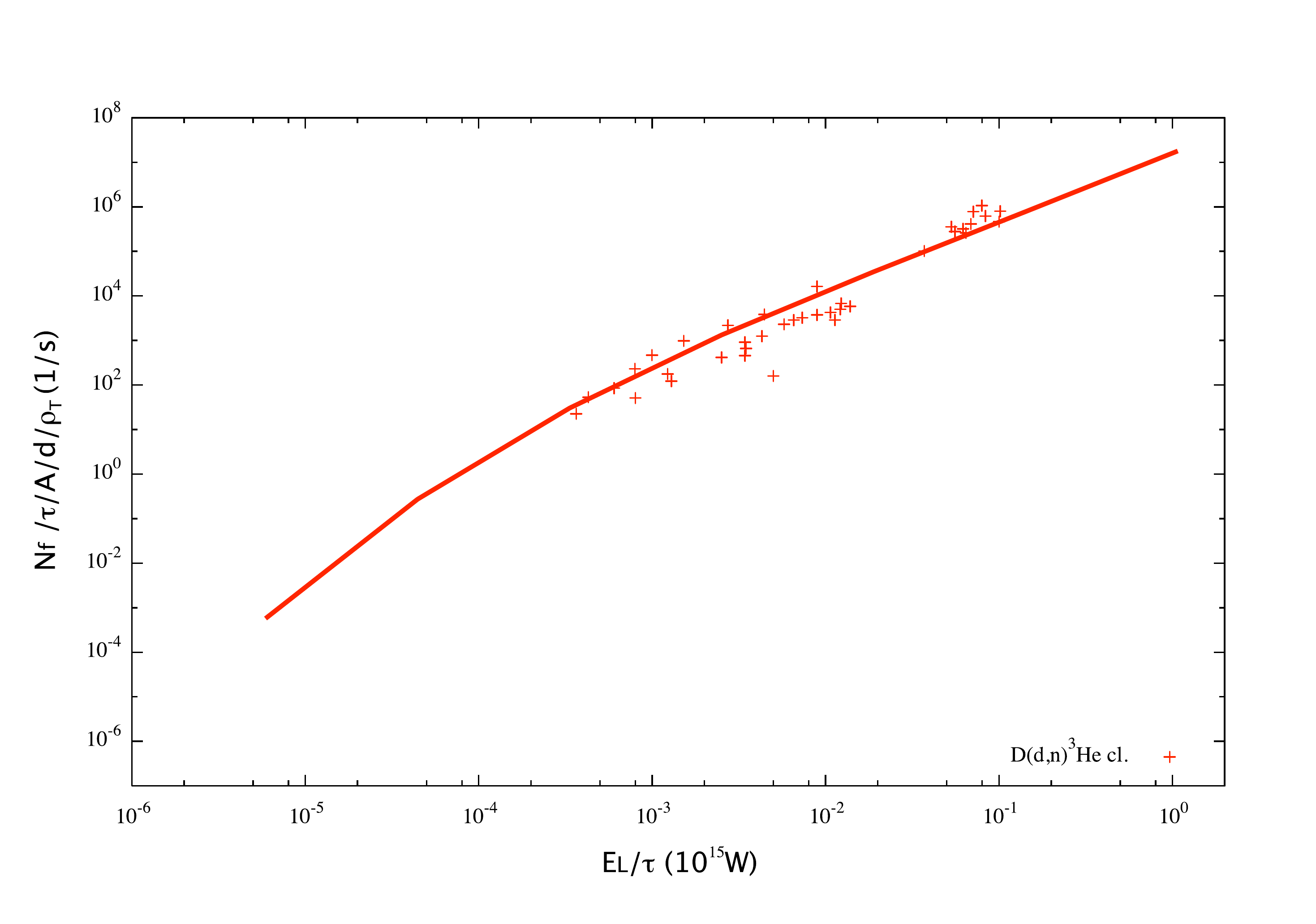}
  \caption{ \label{fig:T019} Same with the bottom panel of Fig.~\ref{fig:RRdd3} but for  the reaction D($d,n$)$^3$He induced 
  by the D$_2$ cluster gas target irradiation. 
  The laser power dependence of $\langle \sigma v d \rangle N_i$ derived with 
  $kT\sim 0.015 \times (E_L/\tau)^{0.19}$ (MeV) (Eq.~(\ref{eq:kt3}))
  for the reaction D($d,n$)$^3$He~(solid line) with the
  D$_2$ cluster gas target.  The experimental data of $\Pi$ are shown by pluses. }  
\end{figure}

In the case of laser-irradiation on the deuterated cluster target, the power law dependence of the 
deuteron plasma temperature on the laser power is expressed by Eq.~(\ref{eq:kt3}).  
We evaluate Eq.~(\ref{eq:elt1}) numerically by using Eq.~(\ref{eq:kt3}) and it is shown in Fig.~\ref{fig:T019} by the solid curve
together with $\Pi$ of the experimental data (shown by pluses).
The laser power dependences of $\langle \sigma v d \rangle N_i$ derived by
assuming the linear dependence of the plasma density on the laser power (Eq.~(\ref{eq:elt1}))
is again in excellent agreement with $\Pi$ of the experimental data using deuterated cluster targets. 

It deserves special mention that 
the derivation of the scaling relation relies on the fact that energy spectra of accelerated ions
are characterized by a Maxwellian-like shape with a temperature. 
One should verify the shape of the spectra of the accelerated ions, before 
extending the result of the scaling law to the laser power beyond PW ($E_L/\tau >$ 10$^{16}$ W).

\section{Summary}
We have derived scaling laws in the fusion yields resulting from intense laser 
irradiations on either solid or cluster gas target, as a function of laser parameters. 
The origin of the power scaling has been studied and it is attributed to the laser power dependence 
of three terms: the reaction rate, the density of the plasma and the projected range of the
plasma particle in the medium.
The density of the plasma is derived to scale linearly as a function of the laser power. 
The resulting scaling relations
can be used, e.g., to estimate the yield of positron emitters produced by deuteron-induced reactions
by means of laser-accelerated deuterons.  
The obtained scaling law enables estimating the fusion yield for a nuclear reaction which has not been investigated 
by means of the laser accelerated ion beams.   

\appendix
\section{Maximum proton energy and plasma temperature}
\label{sec:mpe_pt}
The maximum proton energy of laser-accelerated protons from thin foil targets is 
known to scale as a function of the laser power~\cite{maxpresc}: 
\begin{equation}
\label{eq:pmax}
E_{pmax}=E_{\infty} \tanh^2 \left( \frac{\tau}{2\tau_0} \right),
\end{equation}
where $E_{\infty}$ and $\tau_0$ are 
the potential barrier in which the hot electrons are confined 
and the time in which a proton remains in the vicinity of the accelerating surface charge;
$\tau$ is the pulse duration as before.
$E_{\infty}$ and $\tau_0$ can be written as a function of the laser power:
\begin{equation}
\label{eq:einf}
E_{\infty}=2 m_e c^2 \sqrt{\frac{\eta E_L/\tau}{8.7 (GW)}},
\end{equation}
where $m_e$, $c$ and $\eta$ are the electron mass, the speed of light and the conversion efficiency of the laser pulse into the hot electrons, respectively, and  
\begin{equation}
\label{eq:tau0}
\tau_0\sim \sqrt{\frac{m_p}{2E_{\infty}}},
\end{equation}
where $m_p$ is the proton mass. 
Substituting the equations~(\ref{eq:einf}) and~(\ref{eq:tau0}) into Eq.~(\ref{eq:pmax}), we obtain the laser pulse energy dependence of the maximum proton energy. 
The isothermal fluid model gives the relation between the maximum proton energy and the hot electron temperature~\cite{fuchs_nature,mora}. 
Taking into account this relation, we assume that the plasma temperature is proportional to the maximum proton energy. 
\begin{equation}
\label{eq:kTElap1}
kT\sim \sqrt{\frac{E_L}{\tau}} \tanh^2 \left( \frac{c\tau}{R}\sqrt{\frac{m_e}{m_p}} \left(\frac{\eta}{8.7 \times 10^{-6}} \frac{E_L}{\tau} \right)^{0.25}\right).
\end{equation}

\section{Reaction rate per pair of particles}
\label{sec:rrppA}
\begin{figure}
  \includegraphics[height=.38\textheight, bb=0  0 846 594]{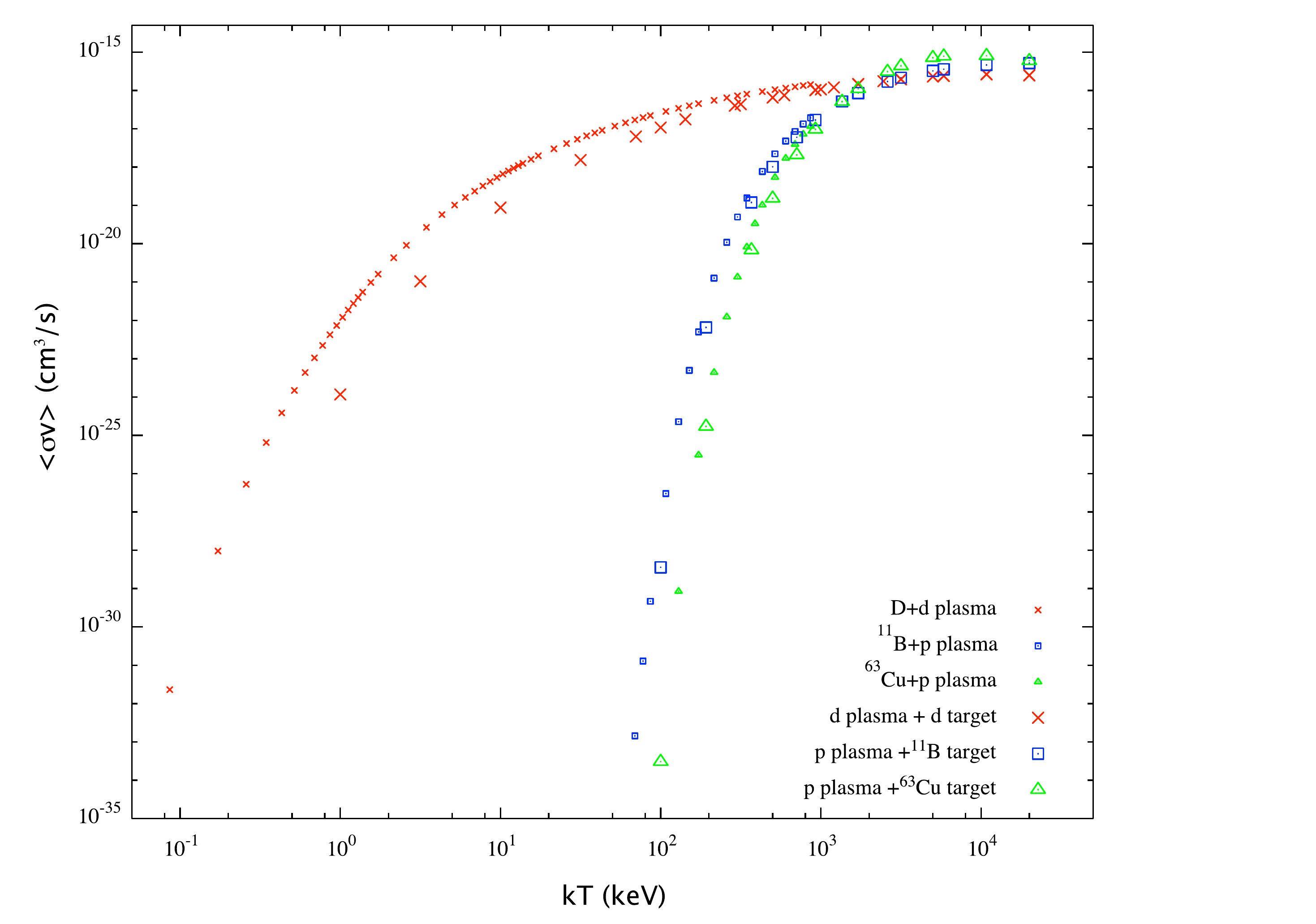}

  \caption{ \label{fig:RR} Reaction rate per pair of particles for  D($d,n$)$^3$He~(crosses), $^{11}$B($p,n$)$^{11}$C~(squares) and 
  $^{63}$Cu($p,n$)$^{63}$Zn~(triangles). The smaller symbols are for the thermonuclear reaction rates~[44,52]. The bigger symbols are for the reaction rate between 
  an ion in the plasma at the specified temperature and an ion at rest.  }
\end{figure}

The thermonuclear reaction rate is evaluated by~\cite{clayton}
\begin{equation}
  \label{eq:sigmav}
  \langle \sigma v \rangle=\int \sigma(v)v \phi(v)dv^3,
\end{equation}
where $\phi(v)$ is the relative velocity spectrum of a pair of ions and it is given by  
a Maxwellian-distribution at the temperature $kT$:
\begin{equation}
 \phi(v)=\left(\frac{\mu}{2\pi kT} \right)^{\frac{3}{2}}\exp \left(-\frac{\mu v^2}{2kT} \right), \label{eq:vs} 
\end{equation}
denoting the reduced mass of the colliding ions in MeV as $\mu$. 
Here we write the reaction cross section as a function of the velocity to keep the consistency in the velocity integral, instead of 
as a function of the incident energy.
Since we consider 
the collisions between the ions in the plasma and the nuclei in the target (HT),
one should take into account that one of the colliding ions is at rest in the laboratory frame.
Therefore the velocity spectrum Eq.(\ref{eq:vs}) is modified as: 
\begin{equation}
 \phi_{HT}(v)=\left(\frac{m_1}{2\pi kT} \right)^{\frac{3}{2}}\exp \left(-\frac{m_1 v^2}{2kT} \right), \label{eq:vsHT} 
\end{equation}
where $m_1$ is the mass of the ions in plasma, instead of the reduced mass. 
One can define an effective temperature as, 
\begin{equation}
 kT^{eff}=\frac{\mu}{m_1}kT.  \label{eq:effectt}
\end{equation}
Using this, the velocity spectrum Eq.~(\ref{eq:vsHT}) is converted to the energy spectrum. The energy distribution is therefore:
\begin{equation}
\label{eq:psi}
\psi(E) dE=\frac{2}{\sqrt{\pi}}\frac{E}{kT^{eff}}\exp\left( \frac{-E}{kT^{eff}}\right) \frac{dE}{(kT^{eff}E)^{1/2}}.
\end{equation}
The reaction rate for the plasma ions colliding with ions at rest is calculated by 
\begin{equation}
  \langle \sigma v \rangle=\int \sigma(E)\sqrt{\frac{2E}{m_1}} \psi(E)dE. \label{eq:rrE}
\end{equation}
In Fig.~\ref{fig:RR} the reaction rates are shown for D($d,n$)$^3$He~(crosses),  $^{11}$B($p,n$)$^{11}$C~(squares) and 
$^{63}$Cu(p,n)$^{63}$Zn~(triangles), respectively.   The thermonuclear reaction rates for the three reactions are given by the small symbols.
The data are retrieved from the NACRE-compilation~\cite{nacre} for the first two reactions and from  
a database of the Hauser-Feshbach statistical model calculations~\cite{rauscher} for the reaction $^{63}$Cu(p,n)$^{63}$Zn. 
While the reaction rates between an ion in the plasma and an ion at rest
are shown by bigger symbols.  
The quantity of interest is the latter and 
the reaction rates in plasmas (smaller symbols) are shown just for comparison.
The reaction rates for the reactions $^{11}$B($p,n$)$^{11}$C and $^{63}$Cu($p,n$)$^{63}$Zn
show steep rises at the temperature around 100~keV. This is due to the negative Q-values for the two reactions.

\section{Effective Temperature}

In Eq.~(\ref{eq:effectt}), the factor in front of the temperature is the same factor 
which connects the relative energy of two colliding particles ($E$) in the center-of-mass system 
and the relative energy in the laboratory system ($E_{lab}$):
\begin{equation}
E=\frac{\mu}{m_1} E_{lab},
\end{equation}
where $E_{lab}=m_1 v^2/2$.  
It is the same either we work out in the laboratory system or in the center-of-mass system. 
We take the center-of-mass system and multiply the temperature by the conversion factor of the energies,
for convenience, so as that the energy $E$ in 
Eq.~(\ref{eq:psi}) is still the energy in the center-of-mass system. 
By using the effective temperature determined by Eq.~(\ref{eq:effectt}),  the velocity distribution
(Eq.~(\ref{eq:vsHT})) is 
\begin{equation}
 \phi_{HT}(v) dv^3=\left(\frac{\mu}{2\pi kT_{eff}} \right)^{\frac{3}{2}}\exp \left(-\frac{\mu v^2}{2kT_{eff}} \right) 4\pi v^2 dv. \label{eq:vsHT2} 
\end{equation}
This is exactly the same with the velocity distribution of Eq.~(\ref{eq:vs}) but the temperature is replaced by the effective temperature.  
The energy distribution Eq.~(\ref{eq:psi}) is obtained by substitution of 
\begin{equation}
E=\frac{\mu}{2} v^2
\end{equation}
in Eq.~(\ref{eq:vsHT2}).

\bigskip
\noindent 
\section*{Acknowledgement}

This work has been carried out in part by financial support of Texas A\&M University. 
The authors acknowledge gratefully discussions with Prof. T. Ditmire and Prof. J. B. Natowitz. 
One of the authors (S.K.) thanks the financial support to attend 
the 49th Course 'Atoms and Plasmas in Super-Intense Laser fields' to the course director Prof. D. Batani. 


%

\end{document}